\documentclass[preprints,article, accept, moreauthors,pdftex]{Definitions/mdpi}
\firstpage{1} 
\pubvolume{1}
\issuenum{1}
\articlenumber{0}
\pubyear{2022}
\copyrightyear{2022}
\externaleditor{Academic Editor: 
 Firstname Lastname} 
\datereceived{} 
\dateaccepted{} 
\datepublished{} 
\hreflink{https://doi.org/} 

\Title{A Deep Learning Approach for Repairing Missing Activity Labels in Event Logs for Process Mining}

\TitleCitation{A Deep Learning Approach for Repairing Missing Activity Labels in Event Logs for Process Mining}

\Author{Yang 
 Lu *
\orcidA{}, Qifan Chen \orcidB{}, and Simon K. Poon \orcidC{}
}

\AuthorNames{Yang Lu, Qifan Chen and Simon K. Poon}

\AuthorCitation{Lu, Y.; Chen, Q.; Poon, S.K.}

\address[1]{%
School of Computer Science, The University of Sydney, Sydney, NSW 2006 
 Australia; qche8411@uni.sydney.edu.au (Q.C.); simon.poon@sydney.edu.au (S.K.P.)\\
}

\corres{\hangafter=1 \hangindent=1.05em \hspace{-0.82em}Correspondence: yalu8986@uni.sydney.edu.au}

\abstract{Process mining is a relatively new subject that builds a bridge between traditional process modeling and data mining. Process discovery is one of the most critical parts of process mining, which aims at discovering process models automatically from event logs. The performance of existing process discovery algorithms can be affected when there are missing activity labels in event logs. Several methods have been proposed to repair missing activity labels, but their accuracy can drop when a large number of activity labels are missing. In this paper, we propose an LSTM-based prediction model to predict the missing activity labels in event logs. The proposed model takes both the prefix and suffix sequences of the events with missing activity labels as input. Additional attributes of event logs are also utilized to improve the performance. Our evaluation of several publicly available datasets shows that the proposed method performed consistently better than existing methods in terms of repairing missing activity labels in event logs.}

\keyword{process mining; 
 business process management; incomplete event logs;  data quality;  data management} 

\begin{document}


\section{Introduction}
Business process management techniques are widely applied in modern information systems such as financial, production, and hospital systems. Traditionally, process analysts model business processes through knowledge gained from interviews, workshops, or documents \cite{dumas2013fundamentals}. On the one hand, modeling business processes by hand can be cost-ineffective and time-consuming. On the other hand, involving human beings to model processes can introduce unavoidable biases. Thanks to the large-scale deployment of computer systems, enterprise data has become more accessible. Process mining, a relatively new subject, was introduced to fill the gap between data mining and traditional process modeling. The goal of process mining techniques is to discover process insights directly from the data collected from target organizations \cite{wmp}. 

One of the most critical parts of process mining is called process discovery, which aims at discovering a business process model automatically from process data. The datasets used to discover process models are called event logs. Each event log is a collection of traces, and each trace is an ordered sequence of events. Each event contains an activity, timestamp, and other attributes.

Different process discovery algorithms have been proposed in the last decade, and some of them can guarantee the production of accurate process models under certain circumstances \cite{wmp}. However, like other data mining techniques, the analysis results are heavily related to the quality of the input datasets \cite{cai_challenges_2015}. Most existing process discovery algorithms assume the event log to be complete, and they may not be able to discover accurate process models when the input event log contains missing data. Missing data in event logs has been defined as one of the major data quality issues in process mining \cite{SURIADI2017132, jagadeesh2013wanna}. Several methods were proposed in the field of process mining to repair event logs with missing data \cite{rogge2013repairing, xu2019profile, liu2021repairing, sim2019likelihood, song2015heuristic}. However, none of these methods can accurately repair missing activity labels in event logs when a large number of activity labels are missing.

In this paper, we focus on repairing the missing activity labels in event logs. Inspired by recent research papers that successfully applied deep learning methods to predict the next activities in ongoing traces, we propose an LSTM-based prediction model to predict the missing activity labels. The prediction model takes both the prefix and suffix sequences of the events with missing activity labels as input. In addition, additional attributes of event logs are also utilized to improve the performance.

\subsection{Motivation Example}
\begin{figure}[H]
\newlength{\xfigwd}
\includegraphics[width=0.99\textwidth]{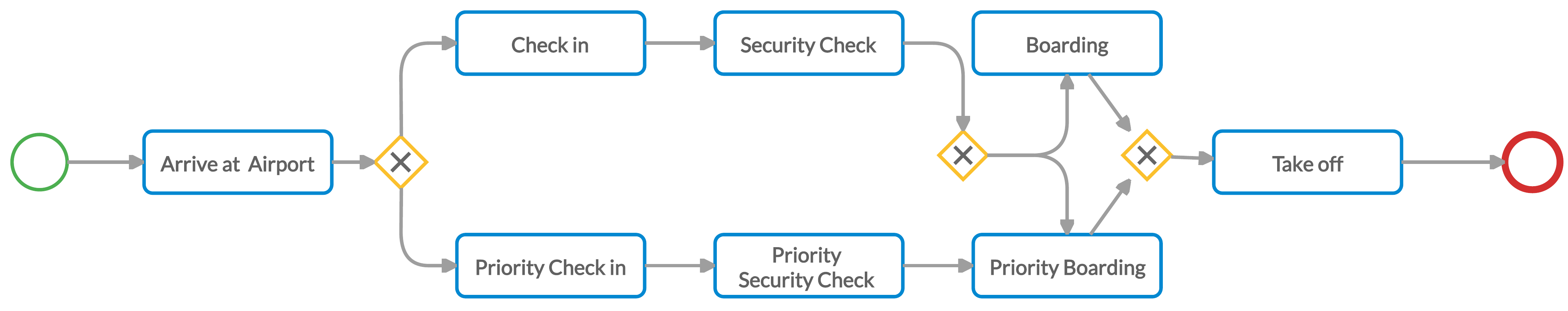}
\caption{Process model discovered from $L_1$.} \label{sample_bpmn_complete}
\end{figure}

Table \ref{example_log_complete} shows an example event log $L_1$, which describes a simple airport process. Each row is an event, which is an execution record of an activity. An event can have multiple attributes. In the example log, each event has an activity label, a resource, and a timestamp. The activity label describes which activity the event recorded, the resource describes the person who performed the event, and the timestamp describes the time when the event was recorded. The event log contains three traces, and each trace is a sequence of events ordered by timestamps. For simplicity, we can write the event log as $L_1$ = \{<Arrive at Airport, Check-in, Security Check, Boarding, Take off>, <Arrive at Airport, Priority Check-in, Priority Security Check, Priority Boarding, Take off>, <Arrive at Airport, Check-in, Security Check, Priority Boarding, Take off>\}. The goal of automated process discovery algorithms is to construct a process model that can accurately describe the process behaviors. For example, if we apply the popular algorithm Split Miner \cite{augusto2019split} on $L_1$, we can obtain the process model as shown in Figure \ref{sample_bpmn_complete}. It is easy to interpret the process model: Some passengers advance through the priority pathways when arriving at the airport, while others advance through the normal pathways. However, passengers with priority tickets can still advance through the normal check-in and security check first, and only advance through the priority boarding in the end. 

\begin{figure}[H]
\includegraphics[width=0.99\textwidth]{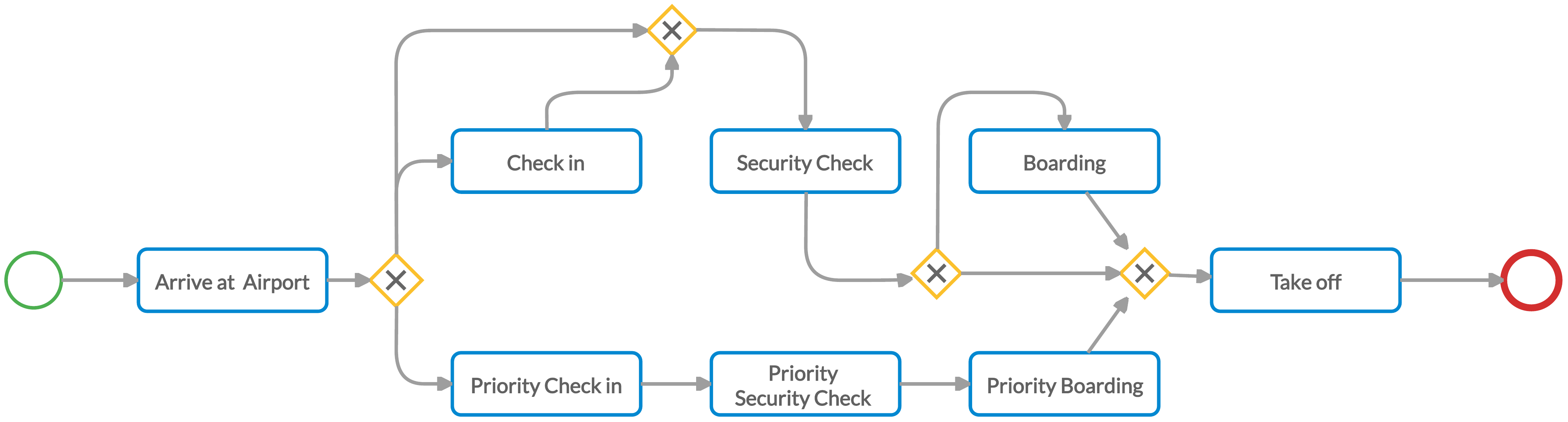}
\caption{Process model discovered from $L_2$. Events with missing activity labels are removed.} \label{sample_bpmn_incomplete_1}
\end{figure}

\begin{figure}[H]
\includegraphics[width=0.99\textwidth]{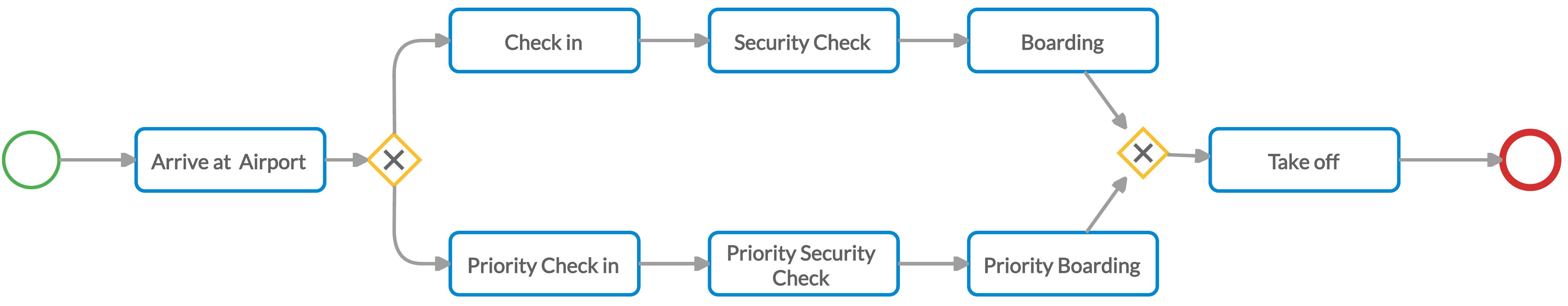}
\caption{A process model discovered from $L_2$. Traces with missing activity labels are removed.} \label{sample_bpmn_incomplete_2}
\end{figure}

Assume there is another event log $L_2$, shown in Table \ref{example_log_incomplete}, which is the same as $L_1$ but with missing activity labels, we can write $L_2$ as $L_2$ = \{<Arrive at Airport, Check-in, Security Check, Boarding, Take off>, <Arrive at Airport, Priority Check-in, Priority Security Check, Priority Boarding, Take off>, <Arrive at Airport, \_, Security Check, \_, Take off>\}. When trying to discover a process model from $L_2$, we can ignore the missing activity labels (Figure \ref{sample_bpmn_incomplete_1}) or the whole traces with missing activity labels (Figure \ref{sample_bpmn_incomplete_2}). None of the process models can accurately describe the process as shown in Figure \ref{sample_bpmn_complete}. For example, in Figure~\ref{sample_bpmn_incomplete_1}, passengers can pass the security check without checking in at the airport. In Figure \ref{sample_bpmn_incomplete_2}, passengers cannot reach  priority boarding after a normal security check.

\begin{table}[H]
\caption{An example event log $L_1$ without missing activity labels.}
\label{example_log_complete}
	\begin{adjustwidth}{-\extralength}{0cm}
		\newcolumntype{C}{>{\centering\arraybackslash}X}
		\begin{tabularx}{\fulllength}{CCCCC}
			\toprule
		\textbf{Event}&\textbf{Trace Id}& \textbf{Activity}&\textbf{Resource}&\textbf{Timestamp}\\
			\midrule
		   $e_{1}$&1&Arrive at Airport& Tom&1/9/2020 12:00:00\\
$e_{2}$&1&Check in& Jack&1/9/2020 12:20:00\\
$e_{3}$&1&Security Check& Thomas&1/9/2020 12:30:00\\
$e_{4}$&1&Boarding& Linda&1/9/2020 14:20:00\\
$e_{5}$&1&Take off& James&1/9/2020 15:00:00\\
\hline
$e_{6}$&2&Arrive at Airport& Alice&2/9/2020 12:10:00\\
$e_{7}$&2&Priority Check in& James&2/9/2020 12:20:00\\
$e_{8}$&2&Priority Security Check& Lucas&2/9/2020 13:30:00\\
$e_{9}$&2&Priority Boarding& Linda&2/9/2020 14:20:00\\
$e_{10}$&2&Take off& Peter&2/9/2020 15:00:00\\
\hline
$e_{11}$&3&Arrive at Airport& Steven&2/9/2020 20:00:00\\
$e_{12}$&3&Check in& Jack&2/9/2020 20:20:00\\
$e_{13}$&3&Security Check& Mark&2/9/2020 20:25:00\\
$e_{14}$&3&Priority Boarding& Linda&2/9/2020 21:00:00\\
$e_{15}$&3&Take off& Ethan&2/9/2020 21:30:00\\
			\bottomrule
		\end{tabularx}
	\end{adjustwidth}

\end{table}

\begin{table}[H]
\caption{An example event log $L_2$ with missing activity labels.}
\label{example_log_incomplete}
	\begin{adjustwidth}{-\extralength}{0cm}
		\newcolumntype{C}{>{\centering\arraybackslash}X}
		\begin{tabularx}{\fulllength}{CCCCC}
			\toprule
		\textbf{Event}&\textbf{Trace Id}& \textbf{Activity}&\textbf{Resource}&\textbf{Timestamp}\\
			\midrule
$e_{1}$&1&Arrive at Airport& Tom&1/9/2020 12:00:00\\
$e_{2}$&1&Check in& Jack&1/9/2020 12:20:00\\
$e_{3}$&1&Security Check& Thomas&1/9/2020 12:30:00\\
$e_{4}$&1&Boarding& Linda&1/9/2020 14:20:00\\
$e_{5}$&1&Take off& James&1/9/2020 15:00:00\\
\hline
$e_{6}$&2&Arrive at Airport& Alice&2/9/2020 12:10:00\\
$e_{7}$&2&Priority Check in& James&2/9/2020 12:20:00\\
$e_{8}$&2&Priority Security Check& Lucas&2/9/2020 13:30:00\\
$e_{9}$&2&Priority Boarding& Linda&2/9/2020 14:20:00\\
$e_{10}$&2&Take off& Peter&2/9/2020 15:00:00\\
\hline
$e_{11}$&3&Arrive at Airport& Steven&2/9/2020 20:00:00\\
$e_{12}$&3&-& Jack&2/9/2020 20:20:00\\
$e_{13}$&3&Security Check& Mark&2/9/2020 20:25:00\\
$e_{14}$&3&-& Linda&2/9/2020 21:00:00\\
$e_{15}$&3&Take off& Ethan&2/9/2020 21:30:00\\
			\bottomrule
		\end{tabularx}
	\end{adjustwidth}

\end{table}

The goal of this paper is to propose a method as a data pre-processing tool that can accurately repair the missing activity labels. The repaired event logs can then be used by process discovery algorithms to discover accurate process models.

\subsection{Contributions of This Paper}
The contributions of this paper include the following:

\begin{itemize}
\item To the best of our knowledge, this is the first paper applying artificial neural networks to predict missing activity labels in event logs for process mining.

\item A LSTM-based artificial neural network to repair missing activity labels in event logs.

\item Experiments on publicly available datasets under various settings show that our method can accurately repair missing activity labels even when a large proportion of activity labels were missing in an event log.

\end{itemize}

The rest of this paper is structured as follows: Section \ref{sec:related_work} is a literature review of related work. Section \ref{sec:preliminaries} introduces some preliminary concepts used in this paper. Our proposed method is presented in Section \ref{sec:method}. Section \ref{sec:evaluation} presents the evaluation results. Finally, our paper is concluded in Section \ref{sec:conclusion}.

\section{Related Work}
\label{sec:related_work}
\subsection{Process Discovery Algorithms}

Various process discovery algorithms were proposed in the last decade. Alpha algorithms were one of the earliest groups of process discovery algorithms that can construct Petri nets automatically based on the event logs. The original version \cite{van2004workflow} of the alpha algorithms can guarantee the discovery of certain behaviors in process models when the input event log is noise-free and can satisfy certain completeness requirements. However, the original algorithm cannot discover accurate process models with complex behaviors. Later research papers extended the original alpha algorithm to discover short loops \cite{de2004process}, invisible tasks \cite{wen2010mining, guo2016mining}, and non-free-choice behaviors \cite{wen2007mining, guo2016mining}. Alpha algorithms produce desirable results on noise-free data, but the performance can be heavily affected when trying to discover process models from real-life event logs. The heuristics miners \cite{weijters2011flexible, weijters2006process, vanden2017fodina} were proposed based on the alpha algorithms to handle noises in event logs.

A challenge for process discovery algorithms was whether the algorithm can guarantee the production of ``sound'' process models \cite{wmp}, which is the precondition for process models to be used for process simulation or conformance checking (i.e., a group of algorithms to check if the given process model conforms to the input data). None of the algorithms stated above can guarantee the production of sound process models. To solve the problem,   inductive miners \cite{leemans2013discovering, leemans2013discoveringi, leemans2014discovering, leemans2018scalable, leemans2016using, leemans2017modeling, lu2020novel} were proposed. Inductive miners always return a process notation called process trees, which can be translated into equivalent block-structured Petri nets \cite{leemans2013discovering}. As a result, inductive miners can always discover sound process models. Similar to the alpha algorithms, inductive miners were also shown to discover certain process behaviors when the input event log is complete. Although inductive miners can guarantee the production of sound process models, the behaviors that can be represented by process trees are limited. Recently, the split miner \cite{augusto2019split} was proposed; it can produce sound process models for most of the time. Instead of process trees, the split miner can discover BPMNs directly. As a result, the split miner can discover more process behaviors.

Besides the process discovery algorithms stated above, there are also other types of algorithms  for discovering process models, such as   genetic algorithms (e.g., \cite{van2005genetic, buijs2014quality}), the ILP algorithm (e.g., \cite{van2008process}), and machine-learning-based algorithms (e.g., \cite{sommers2021process}). However, most of these methods rely on the ordering of events within traces to discover process models. The ordering of events within traces could be incorrect when there are missing activity labels or events \cite{SURIADI2017132}.

In a nutshell, there are various process discovery algorithms to choose from when discovering process models. However, most process discovery algorithms require the event logs to satisfy a certain degree of completeness requirements. The performance of existing process discovery algorithms can be affected if a large number of activity labels in event logs are missing.

\subsection{Missing Data in Event Logs}
There are some research papers focusing on handling missing data in event logs for process mining. In \cite{SURIADI2017132, jagadeesh2013wanna},  missing data was defined as one of the data quality issues for event logs. In \cite{rogge2013repairing}, researchers relied on Generalized Stochastic Petri nets (GSPNs) and Bayesian Network models to repair event logs with missing events. Rogge et al. \cite{rogge2013repairing} was the first paper to address the missing data issue in process mining. However, when a Generalized Stochastic Petri net cannot be derived from the event logs (e.g., when a large number of events are missing), the event log may not be accurately repaired. Similar to \cite{rogge2013repairing}, Song et~al.~\cite{song2015heuristic} also relied on process models to repair missing events.

In PROELR \cite{xu2019profile} and SRBA \cite{liu2021repairing}, researchers firstly applied trace clustering methods to cluster ``complete traces'' (i.e., traces without missing activity labels). Each ``incomplete trace'' (i.e., traces with missing activity labels) would then find the cluster that was closest to it. Finally, the incomplete traces were repaired based on the features of their corresponding trace clusters. Both \cite{xu2019profile, liu2021repairing} require a large amount of ``complete traces'' in event logs. As a result, they cannot handle the case when most traces in event logs contain missing activity labels. Furthermore, the performance   \cite{xu2019profile, liu2021repairing} can drop when the event logs contain a large number of missing activity labels.

The MIEC \cite{sim2019likelihood} is a multiple-imputation-based method to repair missing data in event logs. Besides repairing missing activity labels, it can also repair all other missing attributes in event logs. The MIEC relied on the dependency relations between event attributes. For example, some activities may always happen on weekends or be performed by a specific group of people. It may not be able to effectively repair event logs when such dependency relations do not exist, or the event log contains limited attribute data.

Instead of trying to repair missing data in event logs, Horita et al. \cite{horita2020extraction} applied the decision tree learning algorithm to discover the tendency of missing values in event logs. The output of \cite{horita2020extraction} is a decision tree that indicates the conditions that there is likely to have missing data in event logs (e.g., there is an event with a missing activity label when a certain activity happens before it).

Although a few methods were developed to repair missing data in event logs, a method that is capable of accurately repairing a large amount of missing activity labels is still needed in this field.

\subsection{Next Activity Prediction in Event Logs}
Recently, artificial neural networks have been applied to predict the next events in event logs. The goal of the next activity prediction algorithms is to predict the activity label (or other attributes) of an event in a trace given its prefix sequence. Different neural networks have been proposed to make the prediction as accurate as possible. For example, Tax et al. \cite{tax2017predictive} and Camargo et al. \cite{camargo2019learning} designed LSTM models for next activity prediction. More specifically, Tax et al. \cite{tax2017predictive} applied one-hot vector encoding to encode all categorical variables, while Camargo et al. \cite{camargo2019learning} applied embedding algorithms to obtain vector representations of categorical variables.

To further improve the accuracy of LSTM-based models, Pasquadibisceglie et al. \cite{pasquadibisceglie2021multi} designed a multi-view LSTM based model that took both control-flow information (i.e.,~the ordering of activities in traces) and other event log attributes (e.g., the person who performed each activity) as input for next activity prediction. Lin et al. \cite{lin2019mm} implemented an encoder-decoder structure of LSTMs to predict the next activities. In \cite{lin2019mm}, a customized layer called ``modulator'' was designed to assign different attributes with different weights. Taymouri et al. \cite{taymouri2020predictive} combined generative adversarial nets (GANs) with LSTM models to achieve high-accuracy prediction.

Besides LSTM models, other neural network structures were also designed by researchers for next-activity prediction. For example, Pasquadibisceglie et al. \cite{pasquadibisceglie2019using} converted event logs into 2-D representations and designed a neural network model based on a CNN. A stacked autoencoder-based deep learning approach was designed in \cite{mehdiyev2020novel}.

The methods stated above can achieve high accuracy. However, as their goal is to predict next activities in ongoing traces, only information in the prefix can be used for prediction. When dealing with missing activity labels in event logs, both prefix and suffix sequences can be used.

\section{Preliminaries}
\label{sec:preliminaries}
\subsection{Problem Definition}
In this section, we introduce some basic concepts used in this paper. 

\begin{Definition}[Event log, Trace, Activity, Event]
An event log $L$ is a multiset of traces. A trace $t$, $t \in L$,  is an ordered sequence of events. Assuming $A$ is the set of all possible activities, an event $e$ is an execution record of an activity $a \in A$. $\#_n(e)$ denotes the value of attribute $n$ for event $e$. For example, $\#_{activity}(e)$ refers to the activity label associated with $e$, and $\#_{timestamp}(e)$ refers to the timestamp of event $e$.
\end{Definition}

\begin{Definition}[Missing Activity Label]
The activity label of event $e$ is missing if $\#_{activity}(e) = \_$.
\end{Definition}

For example, for event log $L_2$ in Table \ref{example_log_incomplete}, the activity labels of events $e_{12}$ and $e_{14}$ are missing.

\begin{Definition}[k-Prefix and k-Suffix of an Event]
Suppose a trace $t \in L$ where $t = <e_1, e_2, e_3, ...,$\linebreak  $e_n>$. The k-Prefix of event $e_i$ where $e_i \in t$ is the ordered sequence $<e_{i - k}, e_{i - k + 1}, ..., e_{i - 1}>$, and the k-Suffix of event $e_i$ is the ordered sequence $<e_{i + 1}, e_{i + 2}, ..., e_{i + k}>$. In this paper, when talking about the prefix and suffix sequences, we refer to the activity sequences. For example, the k-Suffix of event $e_i$ refers to the ordered sequence $<\#_{activity}(e_{i + 1}), \#_{activity}(e_{i + 2}), ..., \#_{activity}(e_{i + k})>$.
\end{Definition}

In this paper, we focus on repairing events with missing activity labels within event logs.

\subsection{Long Short Term Memory (LSTM)}
The method we propose in this paper is based on LSTM \cite{hochreiter1997long}, which is a common artificial recurrent neural network structure in the deep learning field. LSTM networks are especially suitable for analyzing time-series data and are resistant to the vanishing gradient problem. As mentioned in Section \ref{sec:related_work}, many LSTM-based artificial neural network structures have been proposed by researchers recently to predict next events in ongoing traces.

The definition of an LSTM unit we applied in this paper is presented in the following equations:
\begin{align}
    \mathbf{f}_{g}^{(t)}&=sigmoid(\mathbf{U}_{f} \mathbf{h}^{(t-1)}+\mathbf{W}_{f} \mathbf{x}^{(t)}+\mathbf{b}_{f}) \label{eq:1}\\ 
    \mathbf{i}_{g}^{(t)}&= sigmoid(\mathbf{U}_{i} \mathbf{h}^{(t-1)}+\mathbf{W}_{i} \mathbf{x}^{(t)}+\mathbf{b}_{i}) \label{eq:2}\\ 
    \mathbf{\Tilde{c}}^{(t)}&= tanh(\mathbf{U}_{g} \mathbf{h}^{(t-1)}+\mathbf{W}_{g} \mathbf{x}^{(t)}+\mathbf{b}_{g}) \label{eq:3}\\ 
    \mathbf{c}^{(t)}&= \mathbf{f}_{g}^{(t)}  \circ \mathbf{c}^{(t-1)}+\mathbf{i}_{g}^{(t)} \circ \mathbf{\Tilde{c}}^{(t)} \label{eq:4}\\ 
    \mathbf{o}_{g}^{(t)}&= sigmoid(\mathbf{U}_{o} \mathbf{h}^{(t-1)}+\mathbf{W}_{o} \mathbf{x}^{(t)}+\mathbf{b}_{o}) \label{eq:5}\\ 
    \mathbf{h}^{(t)}&= \mathbf{o}_{g}^{(t)} \circ tanh(\mathbf{c}^{(t)}) \label{eq:6}\\ 
    &\forall t \in \{1,2,\dots, k\}. \nonumber
\end{align}

In the equations above, $\{\mathbf{U}, \mathbf{W}, \mathbf{b}\}$ are trainable parameters. Each LSTM unit takes a single input vector $\mathbf{{x}^{(t)}}$. The input vector will be passed into different gates that decide how the information will flow into and out of the cell. More specifically, Equation \eqref{eq:1} defines the ``forget gate,'' which determines which part of information from the previous cell state to forget. Equation \eqref{eq:2} defines the "input gate," which controls the new information to be stored into the memory. Equations \eqref{eq:3} and \eqref{eq:4} define how the hidden state from the previous LSTM unit $\mathbf{h}^{(t-1)}$ and the new input $\mathbf{x}^{(t)}$ are used to update the cell $\mathbf{c}^{(t)}$. The output gate defined in \eqref{eq:5} describes how the information of the cell state $\mathbf{c}^{(t)}$ will be used to update the hidden state $\mathbf{h}^{(t)}$, which will be passed to the next LSTM unit or subsequent neural network layers. Finally, Equation \eqref{eq:6} defines how the output gate is used to update the hidden state $\mathbf{h}^{(t)}$.

\section{The Proposed Method}
\label{sec:method}
\subsection{Data Preprocessing}
The core idea of the proposed approach is to use supervised learning approaches to predict the missing activity labels in event logs. In other words, the events without missing activity labels are used to train the prediction model, and the prediction model will then be used to predict the missing activity labels in the event log.

Firstly, we need to split the original event log in order to obtain a training dataset where each sample is labeled. We divide all events in the event log into two sets. The first set $E_{complete}$ contains all events with activity labels, and the second set $E_{missing}$ contains all events with missing activity labels. For example, for the sample log $L_{2}$ shown in Table \ref{example_log_incomplete}, $E_{complete} = \{e_1, e_2, e_3, e_4, e_5, e_6, e_7, e_8, e_9, e_{10}, e_{11}, e_{13}, e_{15}\}$, and $E_{missing} = \{e_{12}, e_{14}\}$. For each event in $E_{missing}$ and $E_{complete}$, we obtain the activity labels of its $k-Prefix$ and $k-Suffix$. For events in $E_{complete}$, we will obtain a training dataset where each sample is labeled. The training dataset will then be used to train a neural network model that will be used to predict the activity labels in $E_{missing}$. Besides activity labels, other attributes of events are also included in our proposed deep learning architecture.

For our example event log $L_{2}$, Table \ref{complete_data_set} and \ref{missing_data_set} present the two datasets constructed from $E_{missing}$ and $E_{complete}$ when $k = 3$. For each event, besides its prefix and suffix activity sequences, its resource is also preserved in the dataset. In addition, a special label ``'Missing'' is assigned to the missing activity labels in the prefix and suffix. Using the label ``Missing'' can let the neural network model know that there is supposed to be an activity label at a specific place. For example, suppose there is a trace $<A, \_, \_>$ with two missing activity labels in an event log where activity C always happens two activities after activity A. If we represent the last event's prefix as $<A, Missing>$, we can easily know the trace is $<A, \_, C>$.
\begin{table}[H]
\caption{The example dataset constructed from $E_{complete}$.}
\label{complete_data_set}
	\begin{adjustwidth}{-\extralength}{0cm}
		\newcolumntype{C}{>{\centering\arraybackslash}X}
		\begin{tabularx}{\fulllength}{CCCCCCCCC}
			\toprule
			\textbf{Event}&\textbf{Resource}&\textbf{Prefix\_1}&\textbf{Prefix\_2}&\textbf{Prefix\_3}&\textbf{Suffix\_1}&\textbf{Suffix\_2}&\textbf{Suffix\_3}&\textbf{Label (Activity)}\\
			\midrule
		    $e_1$&Tom&&&&Boarding&Security Check&Check in&Arrive at Airport\\
            $e_2$&Jack&&&Arrive at Airport&Take off&Boarding&Security Check&Check in\\
            $e_3$&Thomas&&Arrive at Airport&Check in&&Take off&Boarding&Security Check\\
            $e_4$&Linda&Arrive at Airport&Check in&Security Check&&&Take off&Boarding\\
            $e_5$&James&Check in&Security Check&Boarding&&&&Take off\\
            \ldots&\ldots&\ldots&\ldots&\ldots&\ldots&\ldots&\ldots&\ldots\\
            $e_{15}$&Ethan&Missing&Security Check&Missing&&&&Take off\\
			\bottomrule
		\end{tabularx}
	\end{adjustwidth}

\end{table}
\vspace{-12pt}

\begin{table}[H]
\caption{The example dataset constructed from $E_{missing}$.}
\label{missing_data_set}
	\begin{adjustwidth}{-\extralength}{0cm}
		\newcolumntype{C}{>{\centering\arraybackslash}X}
		\begin{tabularx}{\fulllength}{CCCCCCCCC}
			\toprule
			\textbf{Event}&\textbf{Resource}&\textbf{Prefix\_1}&\textbf{Prefix\_2}&\textbf{Prefix\_3}&\textbf{Suffix\_1}&\textbf{Suffix\_2}&\textbf{Suffix\_3}&\textbf{Label (Activity)}\\
			\midrule
		    $e_{12}$&Jack&&&Arrive at Airport&Missing&Security Check&Missing&\_\\
            $e_{14}$&Linda&Arrive at Airport&Missing&Security Check&&&Take off&\_\\
			\bottomrule
		\end{tabularx}
	\end{adjustwidth}

\end{table}

Finally, to feed the dataset into the neural network, the categorical variables have to be transformed into numerical values. As a result, categorical data will be passed into an embedding layer first to be transformed into a vector representation. Depending on the choice of embedding methods, more work could be required to process the data. For example, the categorical values may be required to be represented by non-negative integers \cite{guo2016entity}.

\subsection{The Deep Learning Architecture}

\begin{figure}[H]
\includegraphics[width=\textwidth]{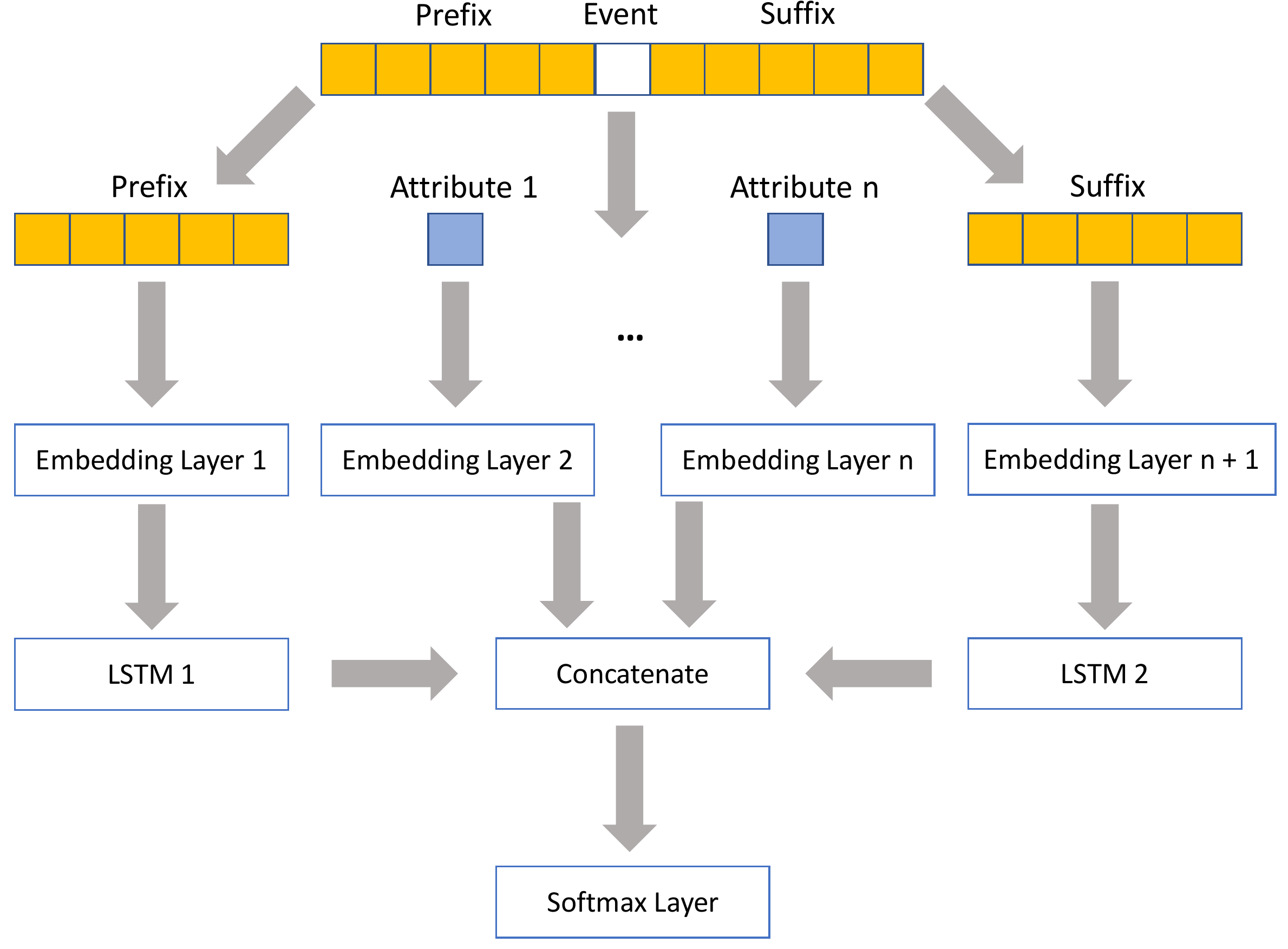}
\caption{Architecture of our proposed neural network.} \label{approach_summary}
\end{figure}

The overall architecture of our proposed neural network is presented in Figure \ref{approach_summary}. The LSTM models for the prefix and suffix sequences are established separately. The deep learning architecture contains two LSTM models. One LSTM model handles the prefix sequence, and the other handles the suffix sequence. 

\begin{figure}[H]

\includegraphics[width=0.75\textwidth]{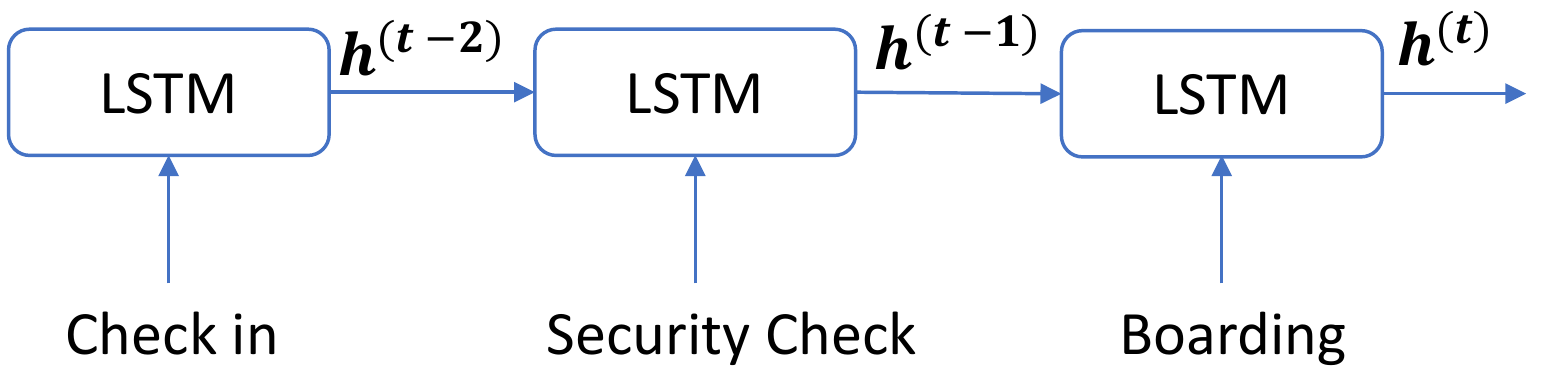}
\caption{Unfolded LSTM network for the prefix of $e_{5}$, the activities are in vector representations.} \label{unfold_lstm}
\end{figure}

For example, Figure \ref{unfold_lstm} shows the unfolded LSTM network for the prefix of $e_{5}$ in $L_{2}$. The LSTM network captures the temporal information for the prefix of $e_{5}$ and outputs a hidden representation $\mathbf{h}^{(t)}$, which is a fixed-size vector. When predicting next activities in ongoing traces, the vector could be passed directly into a dense layer to make the prediction~\cite{tax2017predictive, camargo2019learning}. However, when predicting the missing activity label of an event, we can use the temporal information from both its prefix and suffix. As a result, another LSTM, that captures the temporal information of the suffix sequences of events, is also included in the deep learning architecture.

Besides the temporal information of prefix and suffix sequences, the known attribute values of the events are also used by our proposed method. As shown in Figure \ref{approach_summary}, these attribute values are also passed into embedding layers to be transformed into vector representations. It has to be noted that embedding layers are only needed for categorical variables.

A concatenation layer is used to combine all of the vector representations we obtained. The output vector of the concatenation layer contains the temporal information of the input events' prefix sequences and suffix sequences as well as the information of its attribute values. 

Finally, the output of the concatenation layer is fed into a dense layer that uses the softmax activation function:
\begin{equation}
\text{Softmax}(x_{i}) = \frac{\exp(x_i)}{\sum_j \exp(x_j)} 
\end{equation}

The output of the softmax function is a vector that contains the probability of different activity labels. The activity label with the highest probability will be selected to repair the event. To train the neural network model, the backpropagation algorithm is used to find the optimal trainable parameters. In addition, cross-entropy is used as the loss function.

\section{Evaluation}
\label{sec:evaluation}
To prove that our proposed method can accurately repair missing activity labels in event logs, a large number of experiments were conducted. Overall, we performed two groups of experiments. The first group of experiments compared the performance of our proposed method with existing methods to repair missing activity labels in event logs. The second group of experiments performed further analysis to prove the effectiveness of our method.

We implemented our approach in Python 3.7.1 based on Tensorflow 2.7.0. For the embedding layers, we used the built-in embedding layer in Keras, which requires the categorical variables to be transformed into non-negative integers. In all our experiments, the prefix and suffix lengths were set to five. Zero padding was added if the length of the prefix/suffix was shorter than five (e.g., the event is at the beginning of a trace). In addition, only resources were used as additional attributes to repair missing activities in the experiments of this paper. The dimensions for the embedding layers of the prefix and suffix were set to 100, and the dimension for the embedding layer of resources was set to 16. Probabilistic dropouts of 0.2 were also applied to the outputs of the embedding layers. Moreover, batch normalization was also added to the output vectors of the concatenation layer. Both LSTM networks in our proposed structure contained two layers (32 neurons in the first layers, and 16 neurons in the second layers). During the training process of the model, the training dataset was shuffled first, and 20\% of the training dataset was used as the validation set. To minimize the loss, we used the Nadam optimizer. The maximum number of epochs was 100 (Early stop was set to 10 epochs), the batch size was set to 32, and the learning rate was set to 0.002.

To evaluate the performance of our method, we applied the same evaluation matrix as found in \cite{xu2019profile, liu2021repairing, sim2019likelihood}, i.e., the success rate. It measures the proportion of missing activity labels repaired successfully to the total number of missing activity labels. Equation \eqref{eq:success_rate} defines the success rate, where $m$ is the number of activity labels that are repaired successfully, and $n$ is the total number of missing activity labels.

\begin{equation}
\text{Success Rate} = \frac{m}{n} 
\label{eq:success_rate}
\end{equation}

The experiments utilized several publicly available datasets. In total, our evaluation was based on six publicly available event logs:
\begin{itemize}

\item The Production Process 
 Log\endnote{\url{https://doi.org/10.4121/uuid:68726926-5ac5-4fab-b873-ee76ea412399} (accessed 31 Dec, 2021)}: An event log of a factory's production process. 

\item Hospital Billing\endnote{\url{https://doi.org/10.4121/uuid:76c46b83-c930-4798-a1c9-4be94dfeb741} (accessed 31 Dec, 2021)}: An event log extracted from a regional hospital's ERP system. It contains the processes used to bill bundled packages of medical services.

\item BPI Challenge 2012\endnote{\url{https://doi.org/10.4121/uuid:3926db30-f712-4394-aebc-75976070e91f} (accessed 31 Dec, 2021)}: An event log containing a loan application process in a Dutch financial institute.

\item Sepsis Log\endnote{\url{https://doi.org/10.4121/uuid:915d2bfb-7e84-49ad-a286-dc35f063a460} (accessed 31 Dec, 2021)}: An event log containing processes to deal with sepsis patients in a hospital.

\item Helpdesk\endnote{\url{https://doi.org/10.4121/uuid:0c60edf1-6f83-4e75-9367-4c63b3e9d5bb} (accessed 31 Dec, 2021)}: An event log that describes the ticketing management process in a software company in Italy.

\item BPIC 2013 Incidents\endnote{\url{https://doi.org/10.4121/uuid:500573e6-accc-4b0c-9576-aa5468b10cee} (accessed 31 Dec, 2021)}: An event log of the incident management process in Volvo IT. 

\end{itemize}

\begin{figure}[H]
\includegraphics[width=0.75\textwidth]{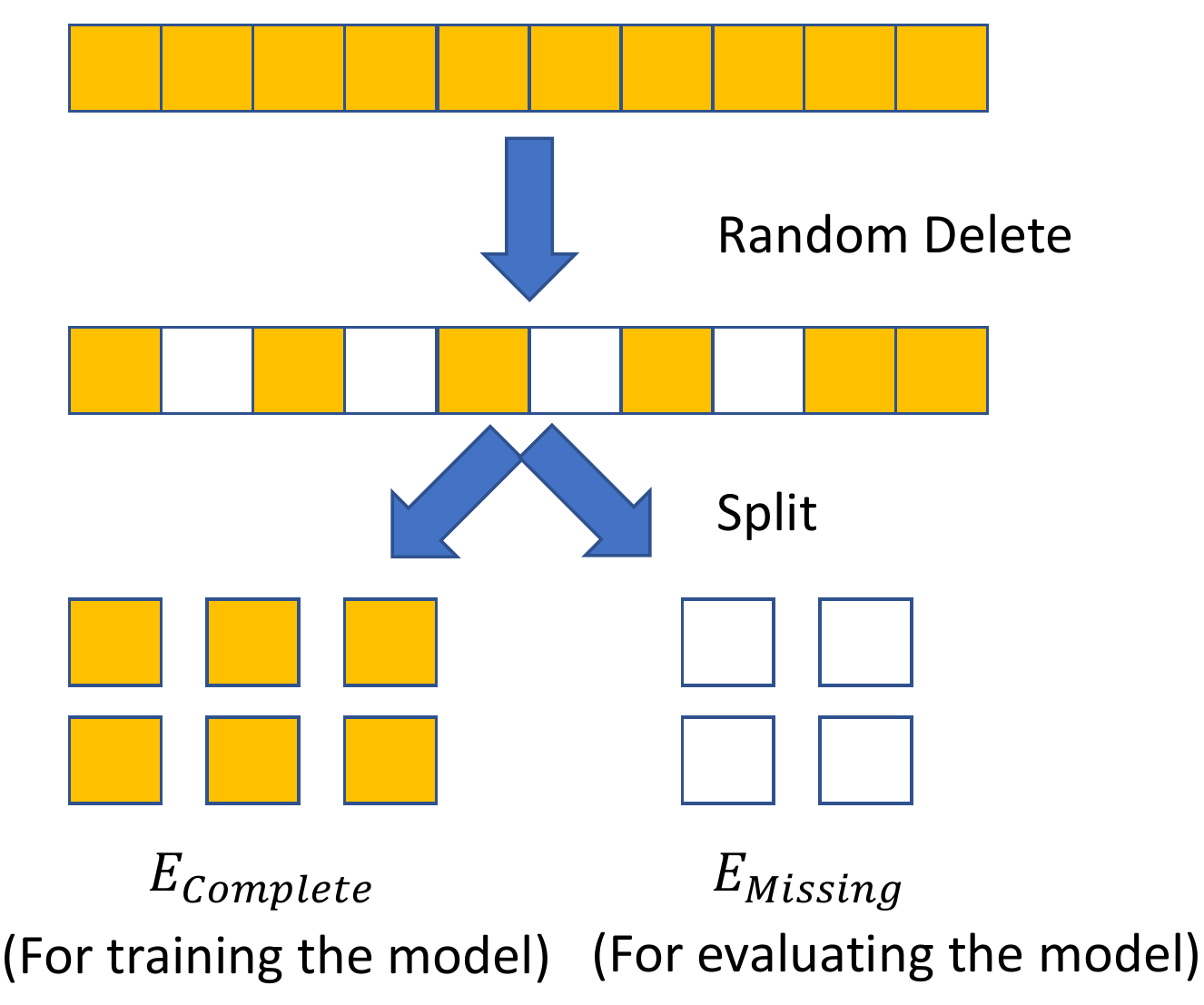}
\caption{Pre-processing publicly available datasets to evaluate our method.} \label{data_prepare}
\end{figure}

The details of all used datasets are presented in Table \ref{meta_data}. For the ``Hospital Billing'' event log, we filtered out all traces with only one or two events. The ``Hospital Billing'' event log was filtered in the same way as in \cite{liu2021repairing}. As shown in Figure \ref{data_prepare}, we firstly randomly deleted a number of activity labels from these event logs. Two datasets were then constructed. The dataset constructed from $E_{complete}$ was used to train the neural network model, and the dataset constructed from $E_{missing}$ was used to evaluate the model and calculate success~rates. 

\vspace{-9pt}

\begin{table}[H]
\caption{Characteristics of the used publicly available datasets.}
\label{meta_data}
	\begin{adjustwidth}{-\extralength}{0cm}
		\newcolumntype{C}{>{\centering\arraybackslash}X}
		\begin{tabularx}{\fulllength}{m{4cm}<{\centering}CCCC}
			\toprule
			\textbf{Dataset}                     & \textbf{Number of Traces} & \textbf{Number of Events} & \textbf{Number of Activities} & \textbf{Number of Resources} \\
			\midrule
		    Production Process Log      & 225      & 4544     & 55           & 31         \\
            Hospital Billing (Filtered) & 69,252    & 412,236   & 18           & 1105       \\
            BPI Chanllenge 2012         & 13,087    & 262,200   & 24           & 69         \\
            Sepsis Log                  & 1050     & 15,214    & 16           & 26         \\
            Helpdesk                    & 4580     & 21,348    & 14           & 22         \\
            BPIC 2013 Incidents           & 7554     & 65,533    & 13           & 1440       \\
			\bottomrule
		\end{tabularx}
	\end{adjustwidth}

\end{table}
\vspace{-6pt}

The settings based on those datasets were slightly different in different experiments, which will be explained in detail in the following subsections.

\subsection{Comparing with Existing Methods}
In the first group of experiments, we compared the performance of our proposed method with \cite{xu2019profile, liu2021repairing, sim2019likelihood}. Since all of these methods were evaluated using publicly available datasets, we can compare our results directly with their results using the same datasets and under the same settings.

We firstly compared our method with PROELR \cite{xu2019profile} and SRBA \cite{liu2021repairing}. In \cite{liu2021repairing}, both methods were evaluated using the ``Hospital Billing Log'' and the ``BPI Challenge 2012'' logs with randomly deleted activity labels. Since both methods rely on trace clustering algorithms, only a small portion of activity labels can be deleted. In addition, only one activity was deleted in each trace.

Following the data preparation methods in \cite{liu2021repairing}, we deleted activity labels of 100, 150, 200, and 300 events from the ``Hospital Billing Log'' and ``BPI Challenge 2012'' logs, respectively, and only one activity label was allowed to be removed from each trace. For each number of missing activity labels, we repeated the same procedure   10 times and reported the average. For example, when removing 100 activity labels from the ``Hospital Billing Log'', we obtained 10 different event logs with 100 missing activity labels, and the success rate reported is the average success rate among the 10 logs.

\begin{figure}[H]
\begin{adjustwidth}{-\extralength}{0cm}
\centering
\includegraphics[width=18cm]{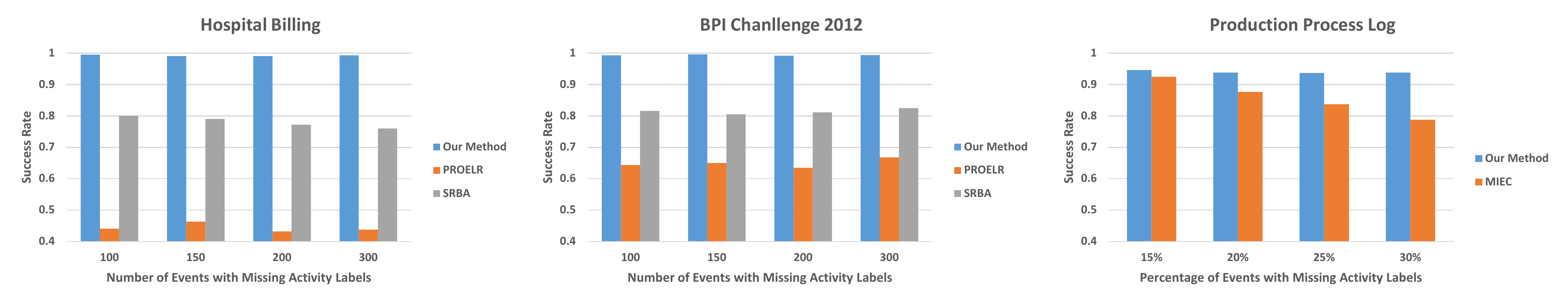}
\end{adjustwidth}
\caption{Comparison of our method with others\label{compare_with_others}.}
\end{figure}

The results for comparing our method with ROELR and SRBA are presented in Table~\ref{compare1} and Figure \ref{compare_with_others}, where the success rates of ROELR and SRBA were referenced directly from~\cite{liu2021repairing}. The success rates of our methods are all above 0.99, which indicates that almost all missing activity labels can be successfully repaired. The success rates are around 0.8 for SRBA and 0.4--0.7 for PROELR, which are lower than our method.

Next, we compared our method with the MIEC \cite{sim2019likelihood}. The MIEC was evaluated by the ``Production Process Log'' in \cite{sim2019likelihood}. To obtain event logs with missing activity labels, the different proportions of activity labels were randomly removed from the log. There were no limits on the number of activity labels deleted in each trace.

Following \cite{sim2019likelihood}, we deleted 15\%, 20\%, 25\%, and 30\% of the activity labels from the "Production Process Log" to obtain event logs with missing activity labels. The evaluation results are presented in Table \ref{compare2} and Figure \ref{compare_with_others}. As in the previous experiment, the results of the MIEC were referenced from \cite{sim2019likelihood} directly, and all the success rates of our method are the average of 10 repeats. Both methods can achieve high success rates when only a small proportion of activity labels are missing in the event log. However, the performance of the MIEC drops when the number of missing activity labels increases. The success rates drop by around 5\% when the number of missing activity labels increases by 5\%, and only 78.8\% of the missing activity labels can be repaired successfully when 30\% of the activity labels are removed. Compared to our method, the success rates remain stable when the number of missing activity labels increases. Around 94\% of the missing activity labels can be repaired successfully at all different levels of missing activity labels.

\begin{table}[H]
\caption{Comparison of our method with PROELR 
 \cite{xu2019profile} and SRBA \cite{liu2021repairing}.}
\label{compare1}
	\begin{adjustwidth}{-\extralength}{0cm}
		\newcolumntype{C}{>{\centering\arraybackslash}X}
		\begin{tabularx}{\fulllength}{Cm{4cm}<{\centering}CCC}
			\toprule
			\textbf{Dataset} & \textbf{Number of Missing Activity Labels} & \textbf{PROELR \cite{xu2019profile}} & \textbf{SRBA \cite{liu2021repairing}.} & \textbf{Our Method} \\
			\midrule
		    \multirow{4}{*}{Hospital Billing}     & 100      & 0.644     & 0.816           & \textbf{0.995} \\
& 150    & 0.650   & 0.805           & \textbf{0.991}       \\
& 200    & 0.635   & 0.811           & \textbf{0.991}         \\
& 300     & 0.668    & 0.825           & \textbf{0.993}         \\
\hline
 \multirow{4}{*}{BPI Chanllenge 2012}     & 100      & 0.441     & 0.800           & \textbf{0.993}         \\
& 150    & 0.463   & 0.790           & \textbf{0.996}       \\
& 200    & 0.432   & 0.772           & \textbf{0.992}         \\
& 300     & 0.438    & 0.760           & \textbf{0.994}         \\
			\bottomrule
		\end{tabularx}
	\end{adjustwidth}

\end{table}
\vspace{-12pt}

\begin{table}[H]
\caption{Comparison of our method with the MIEC
 \cite{sim2019likelihood}.}
\label{compare2}
	\begin{adjustwidth}{-\extralength}{0cm}
		\newcolumntype{C}{>{\centering\arraybackslash}X}
		\begin{tabularx}{\fulllength}{CCCCC}
			\toprule
			\textbf{Dataset} & \textbf{Number of Missing Activity Labels 
} & \textbf{MIEC \cite{sim2019likelihood}} & \textbf{Our Method} \\
			\midrule
		   \multirow{4}{*}{Production Process Log}     
            & 15\% (681)      & 0.925     & \textbf{0.946
}  \\
            & 20\% (908)    & 0.876   & \textbf{0.938}         \\
            & 25\% (1817)   & 0.837   & \textbf{0.937}     \\
            & 30\% (1363)    & 0.788    & \textbf{0.938}       \\
            \bottomrule
		\end{tabularx}
	\end{adjustwidth}

\end{table}

\subsection{Further Analysis of Our Proposed Method}
To further analyze the performance of our proposed method, we evaluated our method with more event logs and missing activity labels. Besides running the experiments on our method, several baselines were also implemented to prove the effectiveness of our method:

\begin{itemize}

\item Prefix Only: A LSTM-based prediction model that predicts the missing activity label of an event using only its prefix sequence.

\item Suffix Only: A LSTM-based prediction model that predicts the missing activity label of an event using only its suffix sequence.

\item Our Method (Without Resources): Our proposed model that uses only prefix and suffix sequences of an event to predict its missing activity label.

\end{itemize}

In total, all six event logs were used to evaluate our method in this section. For each event log, 10\%, 20\%, 30\%, and 40\% of the activity labels were randomly deleted to create missing activity labels. For "Hospital Billing" and "BPI Challenge 2012" logs, instead of deleting a small number of activity labels, a large proportion of activity labels were removed. The evaluation results are presented in Table \ref{all_results} where all success rates were averaged by 10~repeats. The success rates of our method are much higher than using only prefix or suffix sequences to predict the missing activity labels whether the additional attributes (i.e., resources) are used or not. The results prove that, when repairing missing activity labels of events, using the information from both their prefix and suffix sequences can significantly improve the success rates. 

It is also interesting to notice that, except for the production process log, although using additional attributes (i.e., resources) can improve the success rates, applying our method without additional attributes can also obtain high success rates. Our results indicate that our method can be used to repair missing activity labels when additional attributes are not available in event logs. 

\begin{figure}[H]
\begin{adjustwidth}{-\extralength}{0cm}
\centering
\includegraphics[width=18cm]{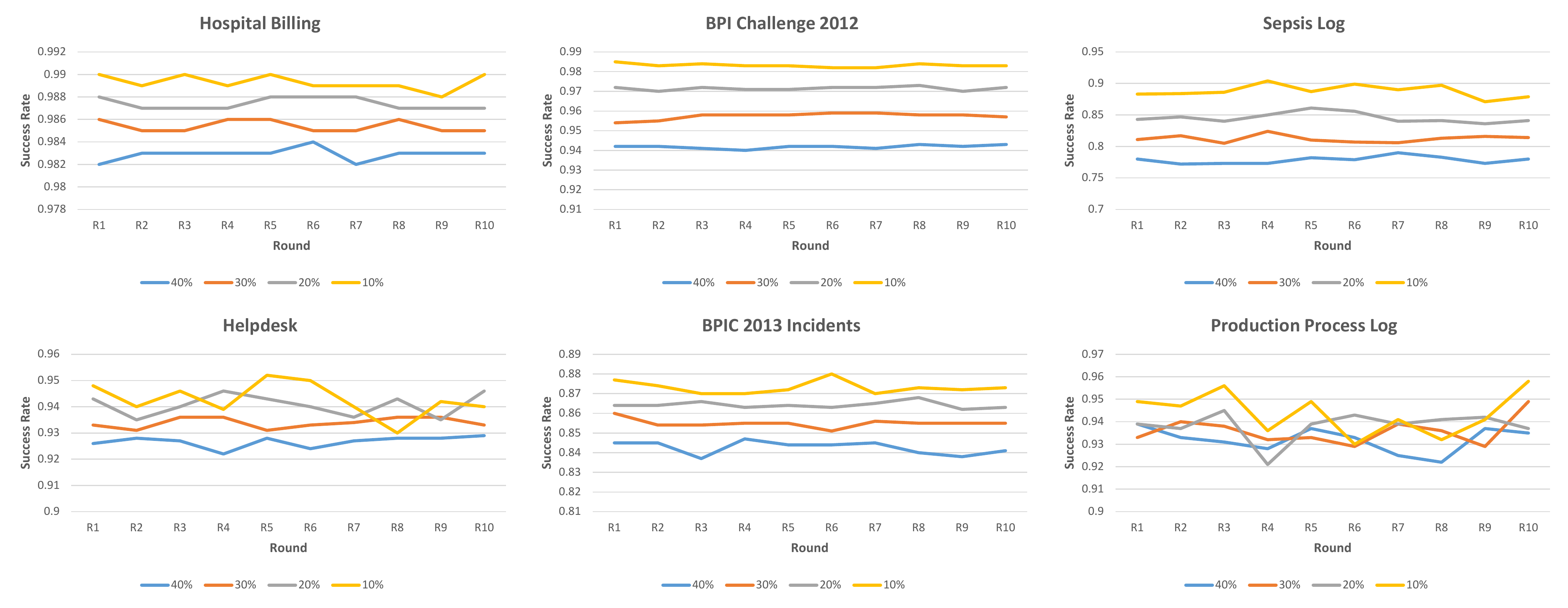}
\end{adjustwidth}
\caption{Successful rates of our method on different datasets in different rounds\label{rounds}.}
\end{figure}

Figure \ref{rounds} shows the success rates of our method each round separately in different event logs when different numbers of activity labels are missing.  Overall, the success rates of our method are stable in different rounds without huge fluctuations. Although the success rates become lower when more activity labels are missing, the drops are slight. On average, the success rate drops by only 0.04 when the number of missing activity labels increases from 10\% to 40\%. These results suggest that our method can accurately repair event logs with a large proportion of missing activity labels.

\begin{table}[H]
\caption{Further analysis of our method.}
\label{all_results}
	\begin{adjustwidth}{-\extralength}{0cm}
		\newcolumntype{C}{>{\centering\arraybackslash}X}
		\begin{tabularx}{\fulllength}{m{4cm}<{\centering}CCCCC}
			\toprule
		\textbf{Dataset} & \textbf{Number of Missing Activity Labels} & \textbf{Prefix Only} & \textbf{Suffix Only} & \textbf{Our Method (Without Resources)} & \textbf{Our Method}\\
			\midrule
		  \multirow{4}{*}{Hospital Billing}     
& 10\% (41223)  & $0.878 \pm 0.002$ & $0.885 \pm 0.002$ & $0.986 \pm 0.001$ & $\mathbf{0.990 \pm 0.001}$ \\
& 20\% (82447)  & $0.865 \pm 0.003$ & $0.880 \pm 0.002$ & $0.983 \pm 0.002$ & $\mathbf{0.988 \pm 0.001}$ \\
& 30\% (123670) & $0.852 \pm 0.002$ & $0.875 \pm 0.003$ & $0.980 \pm 0.001$ & $\mathbf{0.985 \pm 0.001}$ \\
& 40\% (164894) & $0.836 \pm 0.002$ & $0.868 \pm 0.001$ & $0.976 \pm 0.001$ & $\mathbf{0.983 \pm 0.001}$ \\
\hline
 \multirow{4}{*}{BPI Chanllenge 2012}     
& 10\% (26220)  & $0.828 \pm 0.096$ & $0.809 \pm 0.091$ & $0.975 \pm 0.046$ & $\mathbf{0.983 \pm 0.002}$ \\
& 20\% (52440)  & $0.823 \pm 0.002$ & $0.795 \pm 0.002$ & $0.968 \pm 0.002$ & $\mathbf{0.971 \pm 0.002}$ \\
& 30\% (78660) & $0.802 \pm 0.003$ & $0.768 \pm 0.004$ & $0.952 \pm 0.004$ & $\mathbf{0.957 \pm 0.003}$ \\
& 40\% (104880) & $0.778 \pm 0.002$ & $0.735 \pm 0.004$ & $0.932 \pm 0.002$ & $\mathbf{0.942 \pm 0.002}$ \\
\hline
 \multirow{4}{*}{Sepsis Log}     
& 10\% (1521)  & $0.634 \pm 0.020$ & $0.588 \pm 0.017$ & $0.819 \pm 0.022$ & $\mathbf{0.888 \pm 0.016}$ \\
& 20\% (3042)  & $0.602 \pm 0.016$ & $0.545 \pm 0.013$ & $0.763 \pm 0.009$ & $\mathbf{0.846 \pm 0.016}$ \\
& 30\% (4564) & $0.575 \pm 0.017$ & $0.515 \pm 0.010$ & $0.712 \pm 0.017$ & $\mathbf{0.812 \pm 0.012}$ \\
& 40\% (6085) & $0.547 \pm 0.012$ & $0.481 \pm 0.011$ & $0.660 \pm 0.012$ & $\mathbf{0.779 \pm 0.012}$ \\
\hline
 \multirow{4}{*}{Helpdesk}     
& 10\% (2134)  & $0.822 \pm 0.014$ & $0.868 \pm 0.009$ & $0.937 \pm 0.009$ & $\mathbf{0.943 \pm 0.013}$ \\
& 20\% (4269)  & $0.807 \pm 0.014$ & $0.861 \pm 0.008$ & $0.936 \pm 0.007$ & $\mathbf{0.941 \pm 0.005}$ \\
& 30\% (6404) & $0.795 \pm 0.008$ & $0.853 \pm 0.006$ & $0.929 \pm 0.007$ & $\mathbf{0.934 \pm 0.003}$ \\
& 40\% (8539) & $0.786 \pm 0.005$ & $0.847 \pm 0.008$ & $0.922 \pm 0.005$ & $\mathbf{0.923 \pm 0.005}$ \\
\hline
 \multirow{4}{*}{BPIC 2013 Incidents}     
& 10\% (6553)  & $0.662 \pm 0.009$ & $0.712 \pm 0.007$ & $0.831 \pm 0.006$ & $\mathbf{0.873 \pm 0.007}$ \\
& 20\% (13106)  & $0.650 \pm 0.007$ & $0.703 \pm 0.007$ & $0.823 \pm 0.003$ & $\mathbf{0.864 \pm 0.004}$ \\
& 30\% (19659) & $0.634 \pm 0.005$ & $0.690 \pm 0.005$ & $0.816 \pm 0.006$ & $\mathbf{0.855 \pm 0.005}$ \\
& 40\% (26213) & $0.617 \pm 0.008$ & $0.679 \pm 0.004$ & $0.807 \pm 0.005$ & $\mathbf{0.843 \pm 0.006}$ \\
\hline
 \multirow{4}{*}{Production Process Log}     
& 10\% (454)  & $0.481 \pm 0.041$ & $0.520 \pm 0.053$ & $0.568 \pm 0.045$ & $\mathbf{0.944 \pm 0.014}$ \\
& 20\% (908)  & $0.461 \pm 0.018$ & $0.507 \pm 0.034$ & $0.546 \pm 0.029$ & $\mathbf{0.938 \pm 0.017}$ \\
& 30\% (1363) & $0.430 \pm 0.020$ & $0.478 \pm 0.027$ & $0.514 \pm 0.020$ & $\mathbf{0.936 \pm 0.013}$ \\
& 40\% (1817) & $0.416 \pm 0.025$ & $0.461 \pm 0.022$ & $0.500 \pm 0.025$ & $\mathbf{0.932 \pm 0.010}$ \\
            \bottomrule
		\end{tabularx}
	\end{adjustwidth}

\end{table}
\vspace{-12pt}

\section{Conclusions}
\label{sec:conclusion}
In this paper, we proposed a deep learning method to repair missing activity labels in event logs. The method was inspired by recent research papers in the field of process mining that designed artificial neural network models to predict the next activities in ongoing traces. Different from algorithms that predict next activities, to repair the missing activity labels of events, our method uses both their prefix and suffix sequences. The success rates of our method are much higher compared to existing methods in terms of repairing missing activity labels. Additional attributes in the event log can also be utilized to improve the success rates of our method.

It has to be noted that, like other methods used to repair missing activity labels in event logs, such as those found in \cite{xu2019profile, liu2021repairing, sim2019likelihood}, we assumed that we know the exact locations of the missing values. Although our method cannot be applied directly to event logs when the locations of missing values are unknown, our method can be used together with anomaly detection algorithms, such as \cite{10.1007/978-3-030-85469-0_26}. For example, a missing event may exist between two events when the direct succession relation between two consecutive events is identified as an anomaly.

Future work includes the following aspects: First, besides repairing activity labels, we also plan to expand our method to repair other attributes in the event logs (e.g., resources and timestamps). Second,  besides resources, we also aim at evaluating our method using other additional attributes. Finally, we plan to investigate the feasibility of applying our method in online settings.

\vspace{6pt}
\authorcontributions{Conceptualization and Methodology, Y.L., Q.C., S.K.P.; Development, Y.L.; Validation, Y.L., Q.C., S.K.P.; Writing---original draft preparation: Y.L; Writing---review and editing: Y.L., Q.C., S.K.P.; Supervision: S.K.P. All authors have read and agreed to the published version of the manuscript.
}

\funding{This research received no external funding. 
}

\institutionalreview{Not applicable. 
}

\informedconsent{Not applicable. 

}

\dataavailability{All datasets used to evaluate the proposed method are publicly-available, please refer to notes for links to access the datasets. 
} 


\conflictsofinterest{The authors declare no conflict of interest.} 


\begin{adjustwidth}{-\extralength}{0cm}
\printendnotes[custom]

\reftitle{References}

\end{adjustwidth}
\end{document}